\documentclass[prl,twocolumn,Letters ]{revtex4}
\usepackage{graphicx}
\usepackage{color}
\usepackage{amsmath}
\usepackage{color}
\usepackage{bbm}
\usepackage{dsfont}

\def\ket#1{\mathinner{|{#1}\rangle}}

\def\prjct#1{\mathinner{|{#1}\rangle}\!\!\mathinner{\langle{#1}|}}
\newcommand{\coh}[2]{\mathinner{|{#1}\rangle}\!\!\mathinner{\langle{#2}|}}
\newcommand{\braket}[2]{\langle #1  |#2\rangle}

\def\text#1{\textrm{#1}}
\def\id{\mathds{1}}
\def\E{\mathcal{E}}
\def\eq{\begin{equation}}
\def\eeq{\end{equation}}

\def\A{\text{A}}
\def\B{\text{B}}

\def\D{\mathcal{D}}

\begin{document}

\title{
The size of quantum superpositions 
as measured with ``classical'' detectors\\
}

\date{\today}

\author{Pavel Sekatski}
\author{Nicolas Sangouard}
\author{Nicolas Gisin}
\affiliation{Group of Applied Physics, University of Geneva, CH-1211 Geneva 4, Switzerland}

\begin{abstract}
We propose a criterion which defines whether a superposition of two photonic components is macroscopic. It is based on the ability to discriminate these components with a particular class of ``classical'' detectors, namely a photon number measurement with a resolution coarse-grained by noise. We show how our criterion can be extended to a measure of the size of macroscopic superpositions by quantifying the amount of noise that can be tolerated and taking the distinctness of two Fock states differing by N photons as a reference. After applying our measure to several well-known examples, we demonstrate that the superpositions which meet our criterion are very sensitive to phase fluctuations. This suggests that quantifying the macroscopicity of a superposition state through the distinguishability of its components with ``classical'' detectors is not only a natural measure but also explains why it is difficult to observe superpositions at the macroscopic scale.\\

\end{abstract}
\maketitle

\paragraph{Introduction}
Quantum physics is sometimes presented as a theory of microscopic phenomena only, suggesting that there could be a boundary beyond which quantum laws do not apply. However, there is nothing in quantum physics itself that predicts the existence of such a boundary. So either quantum theory is incomplete, or quantum effects apply at any scale but demand a particular effort to be maintained and revealed. This concern provided strong motivations over the last decades to prove through experiments that macroscopic systems can exhibit quantum effects. The question at issue is how to judge whether a quantum system is macroscopic. \\

Let us set the problem. Take an entangled bipartite state 
\eq\label{state}
\ket{\uparrow}_\A \ket{A}_\B + \ket{\downarrow}_\A \ket{D}_\B
\eeq
where the party A is a qubit and B involves two photonic components. Note that even though the terminology of macroscopic superposition is sometimes used, the qubit A is necessary to fix the components $\ket{A}$ and $\ket{D}$ (up to rotations). Assume that one knows how to reveal the entanglement in (\ref{state}). To call this entanglement macroscopic, one wants the states $\ket{A}$ and $\ket{D}$ to be macroscopically distinct \cite{Leggett80}. But how to tell whether this is the case? So far there is no concensus on what such a criterion should be and there could be a variety of different though related concepts. However, we know what macroscopicity cannot be.\\
 
The notion of macroscopicity cannot be invariant under local unitaries, in strong contrast to entanglement. This is already clear in Schr\" {o}dinger's gedanken experiment where a microscopic state of a photonic mode ($\ket{0}$ or $\ket{1}$) is mapped with a unitary transformation onto the macroscopic state of a cat  ($\ket{\text{Alive}}$ or $\ket{\text{Dead}}$). Another example is a series of C-NOT gates that allows one to map a microscopic superposition of qubit states $\ket{\uparrow}$ and $\ket{\downarrow}$ onto a large GHZ-type superposition of $\ket{\uparrow}^{\otimes N}$ and $\ket{\downarrow}^{\otimes N}$. Therefore, in our quest for a macroscopicity criteria, the local unitary invariance has to be abandoned. Furthermore, finding a physically motivated way to break this invariance is the solution to the problem we are aiming at.\\

Several criteria have been proposed recently to define the notion of macroscopicity \cite{Dur02, Bjork04, Korsbakken07, Marquardt08, Lee11,Frowis12, Nimmrichter13}. Specifically, Korsbakken and co-authors \cite{Korsbakken07} linked the macroscopicity of a superposition state carried by an ensemble of qubits with the ease to distinguish its components when only a few qubits are analyzed. This approach (as the majority of available criteria) relies on the partition of the total Hilbert space into individual particles, and there is no such partition for bosonic system. The criteria that we introduce, recognizes that the entangled state (\ref{state}) is macroscopic if its components $\ket{A}$ and $\ket{D}$ are well distinguishable. It demands that these components can be distinguished in a single shot with classical detectors, i.e. the components lead to very different results when measured with detectors whose limited resolution forbids to resolve microscopic states. Indeed the common sense tells us that a property (the distinctness here) is macroscopic if it is first-hand available for us in observation. This  follows the intuition that there is no need for a microscopic resolution to distinguish the dead and alive components of the Schroedinger cat. More precisely, we focus on photon number measurements coarse-grained by noise, a measurement resolving large photon number differences only. It can distinguish the vacuum from a $N$-photon Fock state $\ket{N}$ (as long as N is larger than the detector's uncertainty) but it is unable to discriminate the vacuum from a single photon. This supports the natural claim that a state (\ref{state}) with Fock states $\ket{A}=\ket{M}$ and $\ket{D}=\ket{M+1}$ is a micro-micro entangled state whereas it corresponds to micro-macro entanglement for $\ket{A}=\ket{M}$ and $\ket{D}=\ket{M+N}$ (when $N \gg 1).$ The choice of the photon number measurement for our criterion is arbitrary to some extend. The whole development could be as-well deployed starting with another observable, leading to another hierarchy of macroscopic states. However, the energy (photon number) is a very particular quantity. It is what the human eye measures, and more importantly, it is the only one that can be measured with a passive device and without phase reference.\\ 

\begin{figure}[ht!]
\includegraphics[width=3cm]{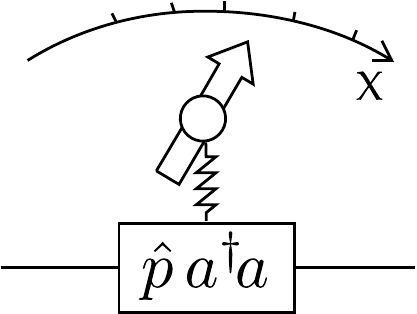}
\caption{A simple model of the photon number measurement.} 
\label{fig1}
\end{figure}

\paragraph{Criterion for macroscopicity} 
A noisy photon number measurement is given by the textbook model depicted in Figure~\ref{fig1}. A classical pointer on a scale $x$ interacts with the state $\ket{S}$ of the mode B and its position is shifted by a value corresponding to the photon number in B \footnote{Physically this may for example correspond to a shift of the kinetic moment of a material pointer given by the radiation pressure Hamiltonian $H = \hat x a^\dag a$.}. The number of photons in $\ket{S}$ is inferred by reading out the final position of the pointer which spans the position $x$ with the probability $p_S(x)=\text{tr}_\B  p_i(x+ a^\dag a) \prjct{S}.$ If the initial position of the pointer $p_i^0(x)$ is $\delta$-peaked around zero, the final probability $p_S^0(x)$ exactly reproduces the statistics of $a^\dag a$ and corresponds to a projective measurement. On the other hand, when the initial position of the pointer has a non-zero Gaussian spread $p_i^\sigma(x)=\frac{1}{\sqrt{2 \pi} \sigma}\exp(-\frac{x^2}{2\sigma^2})$, the probability distribution $p_S^\sigma(x)$ available to the experimentalist does not contain full information on the photon number statistics. Increasing $\sigma$ lowers the resolution of the detector, making it more and more ``classical''. A normally distributed position of the pointer is something one would expect from a classical object, where statistical fluctuations come from a lot of uncorrelated factors \footnote{Remark that if the initial state of the pointer was set to be pure instead, then the detector would perform a weak measurement of the photon number saturating the information/disturbance relation.}.\\

According to our definition, a  macroscopic state (\ref{state}) involves components $\ket{A}$ and $\ket{D}$ that can be distinguished with such a detector. Consider a game where B receives one of these two components (for example prepared by the party A) and has to guess which one has been sent. The probability to make a correct guess in a single shot  
\eq\label{Pguess}
P^\sigma [\ket{A},  \ket{D}]= \frac{1}{2}\Big(1+ D[p_A^\sigma(x),p_D^\sigma(x)]\Big),
\eeq
is related to the trace distance $D[p_A^\sigma(x),p_D^\sigma(x)] = \frac{1}{2}\int dx |p_A^\sigma(x) - p_D^\sigma(x)|$ between the outcome distributions $p_A^\sigma(x)$ and $p_D^\sigma(x)$. The size of the superposition should be related to the amount of noise $\sigma$ that can be tolerated.\\


\paragraph{Quantifying the size}

To define a measure of macroscopicity we have to compare to a reference case for which  there is a natural definition of the size of a superposition. Fock states provide a perfect chance for such a calibration, the probability to guess between two Fock states $ \ket{M}$ and $\ket{M+N}$ with a detector coarse-grained by Gaussian noise is independent of $M$ and reads
\eq\label{Fock}
P_\text{Fock}^\sigma[N] = \frac{1}{2}\Big(1 + \text{Erf}(\frac{1}{2\sqrt{2}}\frac{N}{\sigma}) \Big).
\eeq
The size of the state (\ref{state}) is given by $N$ for which $P_ \text{Fock}^\sigma[N]$ coincides with $P^\sigma[\ket{A},\ket{D}]$.  
However that leaves a free parameter $\sigma$ and to deal with it, we fix the required minimal probability to correctly guess between the two components $P_g$. The maximal tolerable noise $\sigma^{P_g}[\ket{A},\ket{D}]$ is then the solution of the equation
$P^\sigma [\ket{A},  \ket{D}]= P_g$. Given $\sigma^{P_g}[\ket{A},\ket{D}]$, the size of the superposition $(\ref{state})$ 
is obtained by inverting $P_\text{Fock}^{\sigma[\ket{A},\ket{D}]}[N]=P_g$, and corresponds to the $N$ for which the Fock states achieve the same probability $P_g$. If the state $(\ref{state})$ does not reach the required $P_g$, we set its size to be zero. In general the parameter $P_g$ can be suggested by a particular task that one has in mind. For the numerical application, we use $P_g =2/3$ as it is common in the literature on probabilistic algorithms \cite{Aaronson}, in this case \footnote{In general 
\[\text{Size}_{P_g}[\ket{A},\ket{D}] = 2\sqrt{2} \,\text{Erf}^{-1}\Big(2 P_g-1\Big)\, \sigma^{P_g}[\ket{A},\ket{D}]
\]}
\eq\label{sizeP}
\text{Size}_{P_g=\frac{2}{3}}[\ket{A},\ket{D}]  \approx 0.86 \,\sigma[\ket{A},\ket{D}].
\eeq
Remind that by changing the basis at side A, we are free to choose the components $\ket{A'}= \ket{c_\theta A +s_\theta e^{i\varphi}D}$ and $\ket{D'}= \ket{c_\theta D -s_\theta e^{i\varphi} A}$ that maximize the size.\\



\paragraph{Example 1. Optical cat states $\ket{\beta/2}$ and $\ket{-\beta/2}$}
Our first example involves two coherent states with opposite phases $\ket{A} = \ket{-\beta/2}$ and $\ket{D}= \ket{\beta/2}.$ Such states have been at the core of several experiments~\cite{Ourjoumtsev06, Neegaard06, Wakui07}. Obviously $\ket{A} = \ket{-\beta/2}$ and $\ket{D}= \ket{\beta/2}$ have the same energy spectra with the sign information encoded in the phase relation between neighbouring Fock components, and hence they are completely confused by our detector. However it is easy to modify those states to circumvent this problem: Displacing the mode B by $-\beta/2$ brings the components to $\ket{A}=\ket{0}$ and $\ket{D}=\ket{\beta}$ (example 1a) with the corresponding photon number distributions separated by $|\beta|^2$. For fixed $P_g \neq 1$ and large enough $\beta$, the size of this state
\eq
\text{Size}_{P_g}^{(1)}= |\beta|^2 - 2  \Big(\text{Erf}^{-1}(2P_g-1) \Big)^2,
\eeq 
increases linearly with respect to the number of photons, as expected. Remark that the size of the superposition can be increased by further displacing the components to $\ket{A} = \ket{\alpha}$ and $\ket{D} = \ket{\alpha+\beta}$ (example 1b). In the limit $\alpha \gg \beta$ the size of this superposition increases linearly with $\alpha$ and for $|\beta|^2\gg \text{Erf}^{-1}(2 P_g-1)$ the size is proportional to the product $\alpha \,\beta$. The maximal achievable guessing probability is $\lim\limits_{\alpha\to\infty} P^\sigma= \frac{1}{2}(1+\text{Erf} \Big(\frac{\beta}{\sqrt{2}}\Big))$.\\

This example clearly shows that our measure is not invariant under displacement, since the latter is a non-trivial transformation of the energy spectrum. This is well known in the context of homodyne measurements, where the detector noise $\sigma$ can be circumvented by displacing the measured mode. The next example also exploits the non-invariance of the size with respect to displacement.\\

\paragraph{Example 2. Coherent state and displaced single photon}
We recently proposed \cite{Sekatski12} to investigate the quantum features of macro states through a displaced single-photon entangled state $\D(\alpha)_\B\Big( \ket{1_\A, 0_\B} -  \ket{0_\A, 1_\B} \Big)$ which can rewritten as $
\D(\alpha)_\B\Big( \ket{+_\A,-_\B}- \ket{-_\A,+_\B}\Big)$
with $\ket{+}= \ket{0+1}$ and $\ket{-}= \ket{0-1}$ (see \cite{Bruno13, Lvovsky13} for the corresponding experiments). The photon number distributions for $\ket{A}= \D(\alpha)\ket{+}$ and $\ket{D}=\D(\alpha)\ket{-}$ are both of width $\alpha$ and have their means separated by $2\alpha$. For large enough $|\alpha|^2 (>50)$ the statistic of a coherent state follows a normal distribution, 
the guessing probability is a monotonous function of the ratio $\frac{ \sigma}{\alpha}$ (contrary to $\frac{\sigma}{N}$ for the Fock states),
with $\lim\limits_{\alpha\to\infty} P^\sigma \approx 0.899$. Consequently, the size of this state scales as the square root of the photon number and is precisely given by
\eq
\text{Size}_{P_g}^{(2)}=2\, \alpha \, \text{Erf}^{-1}(2 P_g -1) \sqrt{\frac{1}{\pi (2 P_g -1)^2}-2}.
\eeq \\

\paragraph{Example 3. GHZ-like state with overlapping components.}
Let us now focus on the state studied in \cite{Dur02} where the components $\ket{A}=\ket{\phi_1}^{\otimes N}$ and $\ket{D}= \ket{\phi_2}^{\otimes N}$ contain $N$ copies of two non-orthogonal states $|\braket{\phi_1}{\phi_2}|^2=1-\epsilon^2,$ each copy corresponding to a two-level system. Although those states are not photonic but describe spin ensembles, it is easy to generalize our criteria to this case. To do so, replace the number of photons by the population in the excited states (number of $\ket{e}$s) in the definition of the classical detector. 
It is then clear that the size of the superposition depends not only on the relative angle $\epsilon$  between $\ket{\phi_1}$ and $\ket{\phi_2}$ but also on their azimuthal angle.
 For $\ket{\phi_j}=\cos(\theta_j)\ket{g}+\sin(\theta_j)\ket{e}$ with $\theta = \frac{\pi}{4} + (-1)^j \frac{\delta}{2}$ and $\sin(\delta)=\epsilon$, the mean populations of $\ket{A}$ and $\ket{D}$ are maximally separated and for large $N$
\eq
\text{Size}_{P_g}^{(3)}= N \epsilon \sqrt{1- \frac{2 \Big(\text{Erf}^{-1}(2P_g-1)\Big)^2)}{N \frac{\epsilon^2}{1-\epsilon^2}}}
\eeq
which tends to $N \epsilon$ in the asymptotic limit.\\

\begin{figure}[ht!]
\includegraphics[width=7.5 cm]{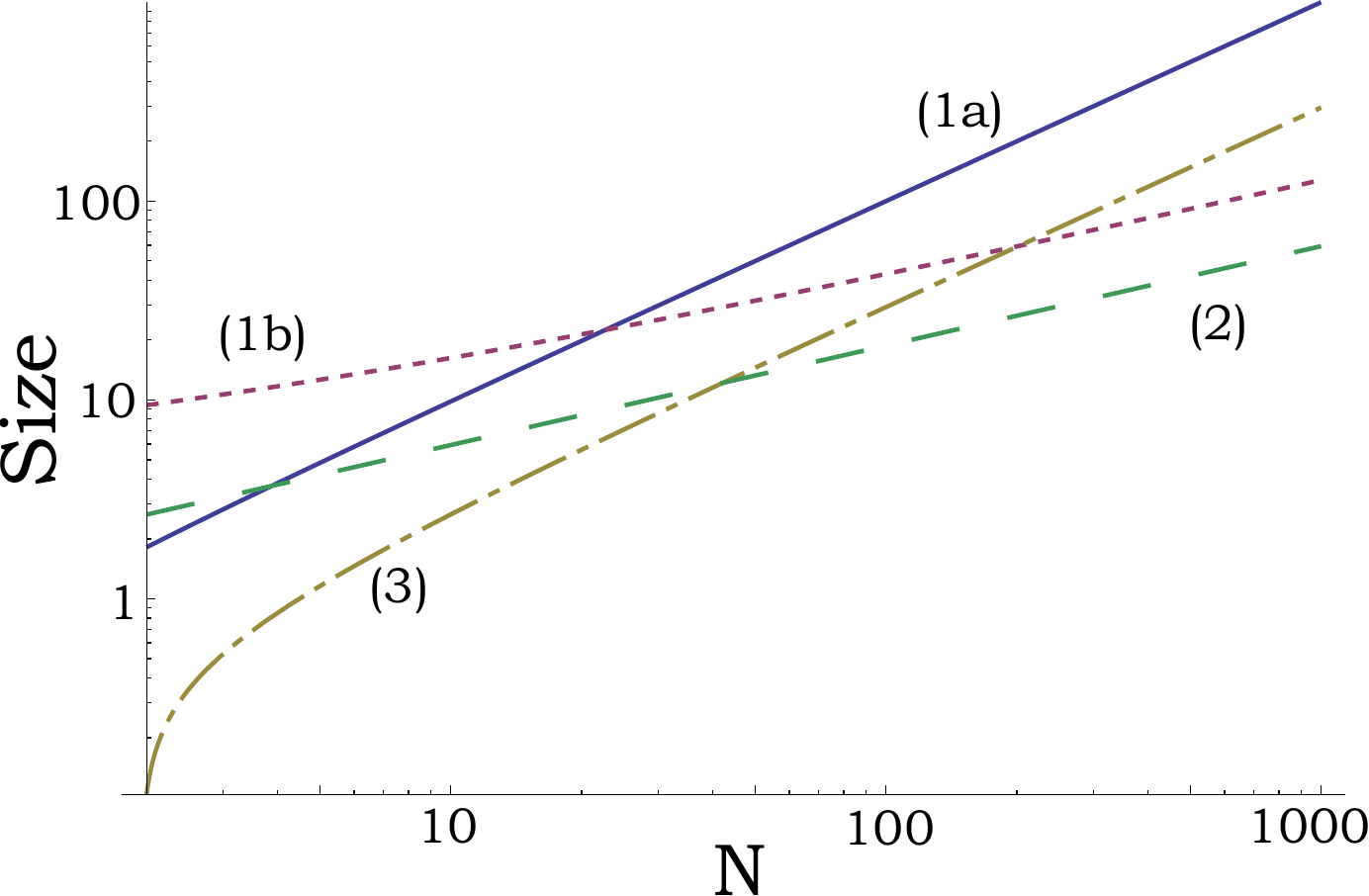}
\caption{Size of several states for $P_g=2/3$ for increasing ``particle number'' $N$. \textbf{(1a)} $\ket{0}$ and $\ket{\beta}$ with $N=|\beta|^2$. \textbf{(1b)} $\ket{\alpha} \text{ and } \D(\alpha)\ket{\beta}$ with $N = |\alpha|^2$ and $|\beta|^2=4$.  \textbf{(2)} $\D(\alpha)\ket{+}  \text{ and } \D(\alpha)\ket{-}$ with $N=|\alpha|^2$. \textbf{(3)} $\ket{\phi_1}^{\otimes N} \text{ and } \ket{\phi_2}^{\otimes N}$ with $\delta=0.3$.} 
\label{fig2}
\end{figure}

\begin{figure}[ht!]
\includegraphics[width=7.5 cm]{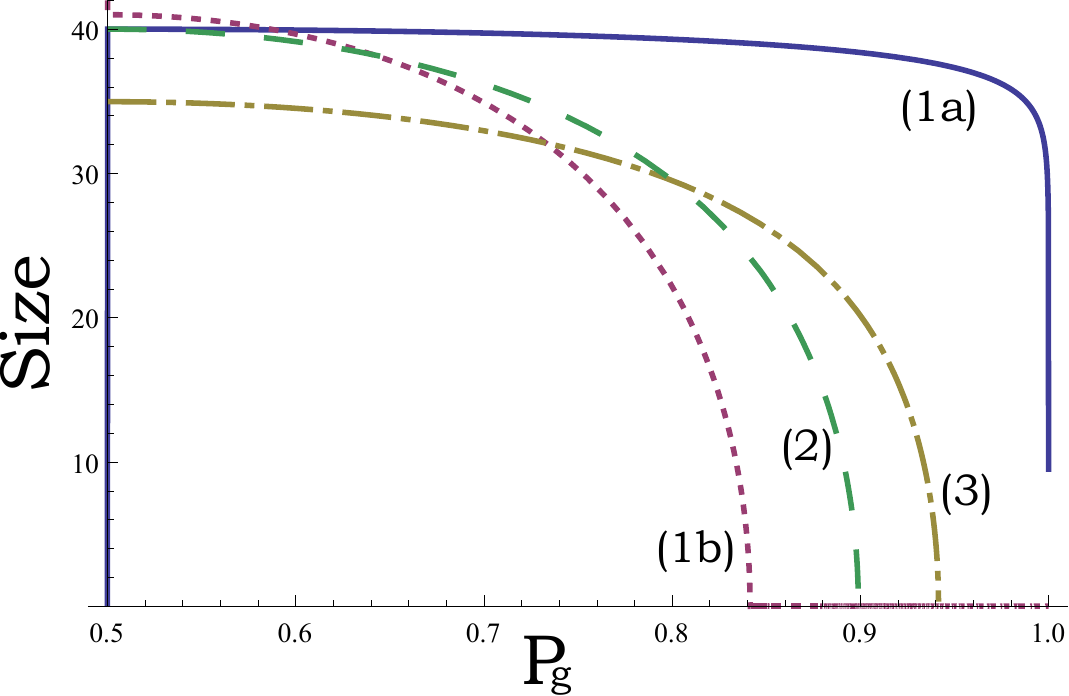}
\caption{Size of several states for a fixed ``particle number'' $N$ as a function of the guessing probability $P_g$.
\textbf{(1a)} $\ket{0}$ and $\ket{\beta}$ with $|\beta|^2=40$. \textbf{(1b)} $\ket{\alpha}$ and $\D(\alpha)\ket{\beta}$ with $|\alpha|^2=400$ and $|\beta|^2=1$. \textbf{(2)} $\D(\alpha)\ket{+}  \text{ and } \D(\alpha)\ket{-}$ with $|\alpha|^2= 400$.  \textbf{(3)} $\ket{\phi_1}^{\otimes N}$ and $\ket{\phi_2}^{\otimes N}$ with $N=500$ and $\delta=0.07$.
}
\label{fig3}
\end{figure}

Fig. \ref{fig2} shows the size of the states that we considered for the guessing probability $P_g=2/3.$ The typical behavior of the size as a function of $P_g$ is given in Fig. \ref{fig3}.\\



\paragraph{A comment on the size for several copies}

An interesting question is how the size of the superposition in our definition behaves when several copies of the state are available. Can one predict what happens when two copies of the states are provided, going from $\{\ket{A}, \ket{D}\}$ to $\{\ket{A}\ket{A},\ket{D}\ket{D}\}$? Unfortunately this is impossible with any definition based on the guessing probability $P_g$ governed by the trace distance. The problem appears already on the classical level: when two copies are measured they give a couple of outcomes $(x,y)$  spanned by $p_A(x) p_A(y)$ or $p_D(x) p_D(y)$. But for the trace distance there exist no general relation between $D[p_A, p_D]$ and $D[p_A^{\otimes 2}, p_D^{\otimes 2}]$, the optimal partition of the outcome plane $(x,y)$ depends on the particular shape of distributions $p_A$ and $p_D$. A good example illustrating this, is the task to guess between two biased coins with face/tail probabilities $p_A= \{p, 1-p\}$ and $p_D=\{1-p,p\}$. One easily verifies that the probability to make a correct guess does not increase after the second throw. Remark that the fidelity between two distributions $F[p_A, p_D]= \int dx \sqrt{p_A(x)p_D(x)} $ behaves nicely with respect to the number of copies $F[p_A^{\otimes N}, p_D^{\otimes N}] = F^N[p_A, p_D]$. But it does not have a nice interpretation in terms of the probability to discriminate between the two states, so a fidelity based definition of the size is not physically motivated, contrary to (\ref{sizeP}).\\

\paragraph{Phase resolution and entanglement}

So far we presented an approach to determine whether the components on side B of the state (\ref{state}) are macoscopically distinct, assuming all the way that they are in a superposition. A certified way to ensure that this is the case, i.e. that the components $\ket{\uparrow}_\A \ket{A}_\B$ and $\ket{\downarrow}_\A \ket{D}_\B$ are indeed superposed and not mixed, is too reveal entanglement between A and B. To do so, the measurements of the number of photons used for the macroscopicity are not sufficient on their own, but  one also needs at least one measuresment in another basis. In the single mode case any such measurement will imply the use of a local oscillator providing phase information, since the underlying POVM necessarily involves coherences between different Fock components $\coh{n}{m}$. In practice, any measurement involving a local oscillator will suffer from a limited phase resolution $\Delta \varphi$. This limitation can be equivalently pictured as a degradation of the local oscillator phase, or as a quantum channel injecting a random phase in the system
\eq\label{phase noise}
\E_{\Delta \varphi}(\rho) = \int d\varphi \,\tilde p(\varphi) \,e^{-i \varphi a^\dag a} \rho\, e^{i \varphi a^\dag a}
\eeq
with a normally distributed random variable $\varphi$ characterized by the standard deviation $\Delta \varphi$. The entanglement in state (\ref{state}) that is experimentally accessible with measurements having a limited phase resolution $\Delta \varphi$ equals the algebraic entanglement in $\E_{\Delta \varphi}^B( \ket{\uparrow}_\A \ket{A}_\B + \ket{\downarrow}_\A \ket{D}_\B)$.\\

The usual phase noise channel (\ref{phase noise}) admits a representation by a unitary evolution of the system plus the environment which is delightful in the present context. Consider an environmental pointer state $\ket{E_0}$ interacting with the system $\rho$ via $U=e^{- i \hat p \,a^\dag a}$. The propagator $U$ shifts the position of the pointer in a controlled way $\ket{E_0(x)}\to\ket{E_0(x-a^\dag a)}$, similarly to the detector defined above. For a pure state $\ket{E_0(x)}$ with a Gaussian envelope and spread $\Delta x$, what we described is nothing else than a weak measurement of the photon number performed by the environment. The state of the system after such interaction is $\rho'=\text{tr}_E\ U \rho \prjct{E_0} U^\dag$. Using $\id_E = \int dp \prjct{p}$ one finds
\eq
\rho'= \int dp \,\tilde p(p) e^{-i p a^\dag a} \,\rho  \, e^{i p a^\dag a} = \E_{\Delta p}(\rho)
\eeq
where $\tilde p (p) = |\braket{p}{E_0}|^2 = |\tilde E_0(p)|^2$ is a Gaussian with standard deviation $\Delta p$. The probability amplitude $\tilde E_0(p)$ in the momenta space is the Fourrier transform of the amplitude $E_0(x)$ in the position space, so the following relation holds $\Delta x = \frac{1}{2 \Delta p}$. Therefore, a standard phase noise channel with fluctuation $\Delta \varphi$ corresponds to a weak photon number measurement of the state by the environment with a pointer of spread $\frac{1}{2 \Delta\varphi}$.\\

The entanglement in the state $\E_{\Delta \varphi}( \ket{\uparrow}_\A \ket{A}_\B + \ket{\downarrow}_\A \ket{D}_\B)$ degrades when the ``which-path'' information ($\ket{A}$ or $\ket{D}$) available to the environment increases. The probability that after the weak measurement the environment correctly guesses between $\ket{A}$ or $\ket{D}$ is lower-bounded by $P^{\frac{1}{2\Delta \varphi}}[\ket{A},\ket{D}]$ (it is not an equality because measuring in the x-basis might not be the optimal choice for the environment). To put it more quantitatively, to experimentally reveal a fraction $E$ of the initial entanglement in the state (\ref{state}), one needs measurements with phase resolution smaller than
\eq
\Delta \varphi = \frac{\sqrt{2} \text{Erf}^{-1}(2P-1)}{\text{Size}_P[\ket{A},\ket{D}]}
\eeq
where $P= \frac{1}{2}(1+\sqrt{1-E^2})$. In other words, for a fixed phase resolution, quantum features are washed out as the size of the state (\ref{state}) increases  -- any trace of entanglement progressively disappears from the measurement results.\\

\paragraph{Conclusion}

We have proposed a measure of the size of macroscopic quantum superpositions. Our criteria relies on the intuition that what makes a property macroscopic is the possibility to observe it with the simplest device. Accordingly, we define two components as being macroscopically distinct if they can be distinguished in a single shot with a noisy photon number measurement (``classical'' detector). The size of a superposition of those components is determined by first quantifying the maximal amount of noise that still allows one to distinguish them with a fixed probability, and then comparing to a superposition of Fock states $\ket{M}$ and $\ket{M+N}$ that we calibrate to be of size $N$. We applied our measure to several examples and extended our criteria to spin ensembles. We further showed that any phase fluctuation can be seen as noisy (weak) measurement of the photon number. Therefore, any single-mode superposition state will only reveal its quantum features under measurements with a phase resolution inversely proportional to the size. Our proposal is thus not only physically motivated, but it also explains why it is so hard to observe quantum features in macro systems. An interesting perspective would be to apply our approach to other detectors that can be reasonably called ``classical'', e.g. detection in the phase space using noisy quadrature measurement. It would also be interesting to extend our approach to states with more than two components $\lambda_0 \ket{0}_\A \ket{A}_\B +\lambda_1 \ket{1}_\A \ket{D}_\B +\lambda_2 \ket{2}_\A \ket{S}_\B + $ etc. There the guessing probability could be replaced by the information obtained by the detector in a single shot. Such a measure would apply to continuous variable entangled states which typically have large Schmidt numbers. \\

\paragraph{Acknowledgements}
We thank B. Sanguinetti and W. D\"ur for stimulating discussions. This work was supported the Swiss SNSF project - CR23I2 127118.

\end{document}